\documentclass[a4paper,12pt]{article}

\begin{document}

\title{The Pioneer anomalous acceleration: \\can we measure the cosmological constant at the scale of the solar system ?}

\author{L. Nottale\\{\small CNRS, LUTH, Observatoire de Paris-Meudon,} \\{\small F-92195 Meudon Cedex, France}}

\date{\today}
\maketitle
\begin{abstract}
An anomalous constant acceleration of $(8.7 \pm 1.3) \times 10^{-8} \; {\rm cm.s}^{-2}$ directed toward the Sun has been discovered by Anderson et al. in the motion of the Pioneer 10/11 and Galileo spacecrafts. In parallel, the WMAP results have definitively established the existence of a cosmological constant $\Lambda=1/ L_{U}^2$, and therefore of an invariant cosmic length-scale $L_{U}=(2.72 \pm 0.10)$ Gpc. We show that the existence of this invariant scale definitively implements Mach's principle in Einstein's theory of general relativity. Then we demonstrate, in the framework of an exact cosmological solution of Einstein's field equations which is valid both locally and globally, that the definition of inertial systems ultimately depends on this length-scale. As a consequence, usual local coordinates are not inertial, so that the motion of a free body of speed $v$ is expected to contain an additional constant acceleration $a_P=v^2/(\sqrt{3} \, L_{U})$,  which is, using the  WMAP five years results, $(6.02 \pm 0.34) \times 10^{-8} \; {\rm cm.s}^{-2}$ when $v\approx c$. Such an effect is too small to contribute significantly to the Pioneer acceleration (since $v_{\rm Pioneer} \approx 12$ km/s $\ll c$), but could be possibly observed in a dedicated space mission.
\end{abstract}


The recent definitive proof of the existence, in Einstein's general relativity equations, of a cosmological constant term $\Omega_{\Lambda}= 0.73 \pm 0.05$ \cite{Perlmutter1998,Riess1998,wmap} (or of an equivalent contribution coming e.g. from vacuum energy) can be considered as a corner stone in the history of cosmology. We shall in this paper investigate one of its possible consequences: namely, its very existence allows the full theory of general relativity to come under Mach's principle, as was initially required by Einstein in his construction. 

The effects of the cosmological constant were up to now considered to hold only at the very large scales. Provided it plays, as we show here, a key role in the determination of inertial systems, it should also manifest itself at local scales. This leads to question whether the Pioneer-Galileo anomalous constant acceleration \cite{anderson98} could result from such an inertial force determined by the cosmological constant. 


Let us first give a short reminder about Mach's principle.
Basing himself on the rotating bucket experiment, in which the existence of a rotational motion can be inferred from the local apparition of inertial forces, Newton concluded to the existence of an absolute space. Two centuries later, Mach proposed another solution, according to which the bucket is in relative motion with respect to the distant bodies of the Universe. Mach's principle was subsequently incorporated as a basic stone in Einstein's construction of general relativity. 

Two levels of Mach's principle were considered by Einstein. The first concerns the nature of inertial systems. The theory of general relativity solves this problem: namely, inertial systems are those whichmove at constant velocity and without rotation relative to the frames in which the universe appears spherically symmetric \cite{weinberg72}.
The second level is the question of the nature and of the amplitude of inertial forces. In a Machian general relativistic framework, they are understood as effects of gravitational induction \cite{einstein17,sciama53,MTW}.

In 1917, Einstein \cite{einstein17} arrived to the conclusion that this second level of Mach's principle would be achieved only provided there exists, at the scale of the universe, a relation between its characteristic length scale $R_U$ and its characteristic mass-energy $M_U$ that reads:
\begin{equation}
\frac{G }{c^2}\;  \frac{ M_U}{R_U}=1.
\end{equation}
A very simple argument has been given by Sciama \cite{sciama53} that allows one to recover fastly this result. In a Machian universe, any body submitted only to gravitation should be considered as free. Therefore its total energy, including its own energy and that of its gravitational coupling with the remaining universe, should be zero. This reads $m c^2 + \sum_i (-Gm \, m_i  /r_i)=0$ in a reference frame where it is at rest, so that one obtains:
\begin{equation}
\frac{G}{c^2}\sum_i \frac{m_i }{r_i}=\frac{G }{c^2}\;  \frac{ M_U}{R_U}=1.
\end{equation} 
Since all solutions based on the cosmological principle are characterized, at the present epoch (well described by dust models), by a conservative relation $M_U=(4/3) \pi \rho_b a^3=$ cst, ($\rho_b$ is the background density and $a=a(t)$ is the scale factor) Einstein reached the conclusion that the Universe had to be static in order implement Mach's principle \cite{einstein17}. This led him to introduce the cosmological constant in the field equations: indeed, in its absence all cosmological solutions are non-static, while its existence allows a unique static solution, the spherical Einstein model.

However, during the twenties the expansion of the Universe was discovered and the Einstein model was found by Eddington to be only metastable. Moreover, in the absence of a cosmological constant, all spherical models reach a maximal radius, which Einstein suggested to identify with the Machian length $R_U$. Therefore Einstein finally concluded that Mach's principle simply led to the constraint that the actual Universe should be closed.

Such a conclusion has been considered unsatisfying, owing to Einstein's initial hope that the whole theory of general relativity be Machian in its essence. Several authors attempted, either to complete general relativity in order to render it Machian, or to unveil its possible hidden Machian structure \cite{barbet82}. 

There is however a simple solution to this question \cite{liwos}, to which the recent measurement of the cosmological constant has given weight. 
Indeed, the cosmological constant is a curvature scalar, and it is therefore the inverse of the square of a length:
\begin{equation}
\Lambda= \frac{1}{L_{U}^2}.
\end{equation}
In standard general relativity, $\Lambda$ is a strict constant, so that the cosmic length $L_{U}$ is an invariant length that is defined at the scale of the Universe. The values of the Hubble constant, $H_0=71 \pm 4$ km/s.Mpc and of the scaled cosmological constant, $\Omega_{\Lambda}=0.73 \pm 0.05$ \cite{wmap}, yield $\Lambda=(1.29 \pm 0.23) \times 10^{-56} {\rm cm}^{-2}$, i.e. $L_{U}= (2.85 \pm 0.25)$ Gpc. With the more recent WMAP 5 years data \cite{WMAP5yr}, one obtains the improved value:
\begin{equation}
L_{U}= (2.72 \pm 0.10)\; {\rm Gpc} .
\end{equation}
Therefore, the mere existence of the cosmological constant allows, whatever the model, to render general relativity Machian, since a universal relation, that reads 
\begin{equation}
\frac{M_{U}}{L_{U}}= \frac{(4/3) \pi \rho_b \, a^3}{L_{U}}= {\rm cst},
\end{equation}
does hold for all dust models.


Let us now address the question of a possible experimental verification of this proposal at the scale of our Solar system. If this point of view is correct, the definition of inertial systems should ultimately be made in relation with the length-scale $L_{U}$, and therefore one should observe inertial forces whose amplitude should be related to the value of the cosmological constant.

Einstein's equations with a cosmological constant read 
\begin{equation}
R_{\mu \nu}-\frac{1}{2} R\;  g_{\mu \nu} - \Lambda g_{\mu \nu}= \chi T_{\mu \nu},
\end{equation}
where $R_{\mu \nu}$ is the Ricci tensor, $R$ the curvature scalar, $\Lambda$ the cosmological constant, $g_{\mu \nu}$ the metrics potentials, $T_{\mu \nu}$ the energy-momentum tensor, and  $\chi=-8 \pi G /c^4$. 
Let us first consider their solution in vacuum around a massive body. It takes the form of Schwarzschild's metric with cosmological constant, namely (we omit the $r^2 d\Omega^2$ term in order to simplify the writing):
\begin{eqnarray}
\label{1276}
ds^2=\left(1-\frac{2m}{r}-\frac{\Lambda}{3} \, r^2 \right) c^2 dT^2 - \left(1-\frac{2m}{r}-\frac{\Lambda}{3} \, r^2 \right)^{-1} dr^2,
\end{eqnarray}
where $m=GM/c^2$. An equivalent form of the Schwarzschild metric is obtained by replacing $dT^2$ by $\psi^2(\tau) d\tau^2$. This generalization will be useful in the following. Note that $r\approx \sqrt{3}\, L_{U}$ is an horizon, since there is an apparent singularity in the metric coefficients when they are written in Schwarzschild coordinates, while it can be suppressed by another choice of coordinates (see e.g. \cite{gautreau1}). Now, the Schwarzschild metric is only a local solution that does not take into account the large scale Universe, while its intervention is necessary if one wants to implement Mach's principle. 


The Universe can be described at large scales by a FRW solution of Einstein's cosmological equations,
\begin{equation}
\label{2764}
ds^2=c^2 dt^2-a^2(t)(dx^2+S^2(x) d\Omega^2),
\end{equation}
where $S(x)= \sin x ,\; x, \; \sinh x $ when the space is respectively spherical, flat and hyperbolic ($k=1,\; 0,\; -1$). The scale factor is solution of an equation of dynamics:
\begin{equation}
\label{4567}
\frac{d^2a}{dt^2}= \left( \frac{\Lambda c^2}{3} - \frac{4 \pi G \rho_b} {3} \right) a.
\end{equation}
Let us introduce the Hubble `constant' $H=\dot{a}/a$ and define the scaled quantities $\Omega_{\Lambda}= \Lambda c^2/3H^2$, $\Omega_M=8 \pi G \rho_b/3 H^2$. We now set:
\begin{equation}
2 \mu= \frac{8 \pi G \rho_b a^3}{3}=\frac{c^3 \; \Omega_M}{H} \left( \frac{k}{\Omega_{\Lambda}+\Omega_M-1} \right)^{3/2}
\end{equation}
which is a constant.
Equation (\ref{4567}) now reads
\begin{equation}
\label{4568}
\frac{d^2a}{dt^2}=  \frac{\Lambda c^2}{3} \: a -  \frac{\mu}{a^2}.
\end{equation}
It is integrated in terms of an energy equation:
\begin{equation}
\label{1298}
\dot{a}^2=  \frac{\Lambda c^2}{3} \, a^2 +  \frac{2 \mu}{a}-k c^2.
\end{equation}
We recognize in the two first terms the cosmological equivalent of the expression that appears in the Schwarzschild metric, ${\Lambda} \, r^2 /3+ 2 G M/(c^2 r)$ , with $r$ replace by $a$ and $M$ by the mass included in a sphere of radius $a$, $M=4 \pi \rho_b a^3 /3$  (in the flat case $k=0$). In an equivalent way, it means that the Hubble constant $H=\dot{a}/a$ is given by 
\begin{equation}
\label{4123}
H^2=  \frac{\Lambda c^2}{3} +  \frac{8 \pi G \rho_b}{3}-á\frac{k c^2}{a^2}.
\end{equation}


None of the two above models can be considered as satisfactory for implementing the Mach principle. Indeed, the Schwarzschild model is locally correct but it fails to incorporate the large scale matter and field distribution which defines the inertial reference frames. The analysis of the Pioneer-Galileo effect has been performed in its framework (without cosmological constant), and it has failed to explain the additional acceleration \cite{anderson02}.

The cosmological model is correct at global scales (provided the cosmological principle be true on very large scales), but it is not at all adapted to scales smaller than our Galaxy radius, and a fortiori at the scale of our solar system. 

Nevertheless, an analysis of the Pioneer-Galileo effect has been performed in its framework by Rosales and Sanchez-Gomez (RS) \cite{RSG99}. Their argument amounts to the following: The inertial reference frames are defined as free falling with respect to the Universe as a whole, so they should be defined with respect to the FRW metric (Eq. \ref{2764}). A free-falling object in such a metric is at rest in a comoving coordinate system, and it is subjected to Hubble's law $v=H_0 d= H_0 c \times t$. The inertial frame therefore accelerates from the center of coordinates, implying an inertial force directed toward it, corresponding to a constant acceleration $H_0 c$.

However, the appearance of an inertial force is instantaneous and it must be determined from the knowledge of the true metric at the point of its occurence. Therefore the RS result is based on the assumption that the FRW metric (and the expansion of the universe) remains valid at the scale of the solar system: namely, they interpret the Pioneer effect as a detection of the cosmological expansion in the solar system. This is a very doubtful hypothesis. Indeed, the energy-momentum tensor that leads to the FRW metric is that of a perfect fluid, whose `particles' are identified with the galaxies: this is a thermodynamical description that becomes wrong at the scales of the fluid `particle' sizes (galaxy radii), and even more inside them. Then there is no reason for the cosmic expansion to apply inside galaxies, and a fortiori at planetary scales that are yet $10^7$ times smaller. This is confirmed by Williams et al., who find no evidence for local scale expansion of the solar system \cite{Williams2004}.


In order to address this problem in a satisfying way, we therefore need an exact solution of Einstein's equation that should be valid both locally and globally. This is the Einstein-Strauss problem, which is solved by a continuous matching between the Schwarzschild line element (Eq.~\ref{1276}) and the cosmological one at the limit of a `vacuole'. Such a matching (including a cosmological constant term) has been performed by Balbinot et al \cite{balbinot88}. It implies a `null apparent mass condition' \cite{eisenstaedt}, according to which the mass retired from the cosmological fluid equals exactly the central Schwarzschild mass. This condition expresses the fact that the background is actually an average description of a large scale distribution of such individual masses. Vacuole models have already been used, for example, in order to find exact solutions of the optical scalar equations, allowing an exact treatment of the gravitational lensing and gravitational redshift and time-delay problems \cite{LN82-84}. This method has in particular put to the light the existence of a new general relativistic factor 2 \cite{DN} in the Rees-Sciama effect, which has been confirmed by subsequent studies using the potential approximation \cite{silk}. 

The matching conditions become particularly simple when the inner and outer metric elements are written in terms of the same coordinate system. We shall therefore perform a change of coordinate system from comoving coordinates to curvature coordinates \cite{gautreau2}. The matching is done on the vacuole limit $r=r_v(\tau)$. The comparison of the angular parts of the metrics implies that $r = a \; S$.

The matching conditions (continuity of the metric potentials and of their derivative) yield the null apparent mass condition, that writes:
\begin{equation}
\label{9783}
M=\frac{4}{3} \pi \rho_b \, r_v^3=\frac{4}{3} \pi \rho_b \, a_v^3 \, S_v^3.
\end{equation}
The cosmic time can be expressed in terms of the new coordinates, namely $t=t(r,\tau)$. The cosmological line element in curvature coordinates is diagonal when \cite{balbinot88}
\begin{equation}
\frac{\partial t}{\partial r}=-\frac{ H r}{1-(H^2+\frac{k}{a^2}) \, r^2 },
\end{equation}
where $a=a[t(r,\tau)]$ and $\dot{a}/a=H[t(r,\tau)]$. The $g_{rr}$ coefficient of the metric then takes the simple form:
\begin{equation}
g_{rr}^{\rm cos}=-\left\{1-\left(H^2+\frac{k}{a^2} \right) \, r^2 \right\}^{-1}.
\end{equation}
Now using Eq. (\ref{4123}), it takes a Schwarzschild-like form:
\begin{equation}
g_{rr}^{\rm cos}=\left( 1-\frac{2\mu S^3}{r}-\frac{\Lambda}{3}\, r^2 \right) ^{-1}.
\end{equation}
The matching with the corresponding coefficient of the Schwarzschild metric is therefore simply performed, since on the matching hypersurface one has $2 \mu S_v^3= 2m$.

The time coefficient of the cosmological metric reads:
\begin{equation}
g_{\tau \tau}^{\rm cos}=\left\{ 1-\left(H^2+\frac{k}{a^2}\right) \, r^2 \right\} \left(\frac{(\partial t /\partial \tau)^2}{1-\frac{k}{a^2}\,  r^2}  \right).
\end{equation}

Up to now the results given were exact. In what follows, we shall now use power expansions up to order 2 in $r$ and $\tau$, which is sufficient for our purpose. Indeed, we consider small time intervals and small distances with respect to the cosmological times and distances. The cosmological scale factor writes to second order:
\begin{equation}
a(t)=a_0 \left(1 + H_0 t -\frac{1}{2} q_0 H_0^2 \, t^2 + ...\right),
\end{equation}
where $q_0= \frac{1}{2} \Omega_M -\Omega_{\Lambda}$. We find for the time transformation the following expansion (here to third order in ($r, \tau$), since we need to know its derivative to second order):
\begin{equation}
t(r,\tau)=\tau + \frac{1}{2} H_0 r^2 + \frac{1}{2} \left(H_0^2 - \frac{\Lambda}{3} \right) r^2 \tau + ....
\end{equation}
Therefore we find 
 $(\partial t  / \partial \tau) ^2=1 + H_0^2 \, (q_0+1) \, r^2$ to the order 2,
so that 
\begin{equation}
\frac{(\partial t /\partial \tau)^2}{1-\frac{k}{a^2}\,  r^2}=1+4 \pi G \rho_b \, r^2 + ... \; .
\end{equation}
Finally, it is found that the outer element reads in curvature coordinates (to the second order as concerns $g_{\tau \tau}$):
\begin{eqnarray}
ds_{\rm cos}^2=\left\{ 1- \left( \frac{\Lambda}{3}-\frac{4}{3} \pi G \rho_b \right) r^2 \right\} c^2 d \tau^2 
-  \left\{1- \left( \frac{\Lambda}{3}+\frac{8}{3} \pi G \rho_b \right)  r^2 \right\}^{-1} dr^2.
\end{eqnarray}
This metric is valid for $r \geq r_v$. Remark that, even in this outer cosmological domain, the $g_{00}$ coefficient is not given, in curvature coordinate, by $1-H_0^2 r^2$ [see Eq. (\ref{4123})]. The matching of the inner and outer line elements is completed by writing the equality of the $g_{\tau \tau}$ coefficients on the hypersurface $r=r_v(\tau)$. One obtains:
\begin{equation}
\psi^2(\tau)=1+4 \pi G \rho_b(\tau) r_v^2(\tau)+ ...=1+\frac{3m}{r_v(\tau)} + ... \;.
\end{equation}
The function $r_v(\tau)$ is solution of the equation $r_v=S_v \, a(r_v,\tau)$ with $S_v=$ cst, where $a(r,\tau)$ is given by
\begin{equation}
a(r,\tau)=1 + H_0 \tau - \frac{1}{2} q_0 H_0^2 \tau^2 -\frac{1}{2}  H_0^2 r^2 +... \; .
\end{equation}
Then $r_v(\tau)=r_v^0 (1+ H_0 \tau)$ to the first order. Therefore we obtain the following inner metric form, valid for $r \leq r_v$: 
\begin{eqnarray}
\label{1627}
ds_{\rm Sch}^2=\left(  1+\frac{3m}{r_v(\tau)}\right) \left(1-\frac{2m}{r}-\frac{\Lambda}{3} \, r^2 \right) c^2 d\tau^2 - \left(1-\frac{2m}{r}-\frac{\Lambda}{3} \, r^2 \right)^{-1} dr^2.
\end{eqnarray}


Two new terms are present in this metric in addition to those which have already been taken into account in the Anderson et al. analysis of the Pioneer effect \cite{anderson02}. The first is the $\tau$ dependence of the $g_{\tau \tau}$ metric potential. It yields a constant acceleration $(3m/r_v) H_0 c$ which is negligible, since $m/r_v \approx 10^{-15}$ for $m=G M_{\odot}/c^2$.

The main new term is the cosmological constant term, which is common to the inner and outer metrics. The mean matter density of the universe (that contributes to the Hubble constant, see Eq.~(\ref{4123})) appears only in the outer metric. In the inner part  of the metric, it is tranformed into the Newtonian potential term of the Sun, whose contribution (with that of planets and satellites), has already been taken into account by Anderson et al. 

However, this form of the metric (Eq. \ref{1627}) cannot yet be directly used for analysing the effect. Indeed, the coordinate $r$ no longer keeps its previous meaning for defining the distance because of the $g_{rr}$ metric coefficient. In agreement with the Pioneer-Galileo measurements, the distance is correctly defined by the travel time of photons. This time should be such that $ds=c \,dt$ for the spacecraft. 

Therefore we shall now jump to an inner coordinate system that is comoving with the spacecraft ($x_{sc}=$ cst, after account of its proper motion). This can be done by applying to the line element (Eq. \ref{1627}) the inverse de Sitter transformation,
\begin{equation}
\label{2085}
r=x \; e^{K t}; \;\; \tau= t- \frac{1}{2K}  \ln \left(1-\frac{\Lambda}{3} x^2 e^{2 K t} \right),
\end{equation}
in which we have set $K=\sqrt{\Lambda/3}=1/R_H$ (the inverse of the horizon radius). This yields (to leading order) the following form of the inner metric:
\begin{eqnarray}
ds_{\rm Schw}^2= \left(1-\frac{2m}{r(x,t)}+\frac{3m}{r_v(x,t)}\right) c^2 dt^2 - \frac{1}{1-\frac{2m}{r(x,t)}} \; e^{2K \, t} \; dx^2,
\end{eqnarray}
where $r(x,t)$ is the function of $x$ and $t$ given in Eq. (\ref{2085}). The next-to-leading order terms give negligible contributions. It is remarkable that this form of the metric, although it is local, includes a cosmological-like expansion term, $e^{2Kt}$ depending on the cosmological constant $\Lambda= 3 K^2$. But it does not include the Hubble constant, which contributes to the expansion only at cosmological scales.

This form of the metric therefore allows us to separate the usual solar system contributions and the cosmological contribution, which is clearly reduced to that of the cosmological constant. Namely, the Anderson et al \cite{anderson98} comparison between the observed motion and the theoretical prediction has accounted for the $(1-2m/r)$ terms in the metric potentials, while there are here two additional contributions:  the $3m/r_v$ term, which yields a negligible additional acceleration, and the cosmological constant term $e^{2Kt}$. 

Therefore, assuming that the various effects can be added in a linear way (to the first approximation), after account of all other effects, the deviation between the observed position of the spacecraft and its position calculated in \cite{anderson98} is in free fall motion in the space-time geometry described by the perturbation metric  
\begin{equation}
ds^2=c^2 dt^2-a_K^2 dx^2,
\end{equation}
with $a_K=e^{Kt}$.\\


We can now relate the effective distance $l=c t_0$ measured on the light cone to the local radial coordinate. We have on the light cone $ds^2=0$, so that the distance $l$ is such that $dl=c\, dt=a_K dx$, while $x=r/a_K$, so that $a_K\, dx= dr-(\dot{a_K}/a_K) r dt$. Then $dl= dr /(1+Kr)$, which is integrated as
\begin{equation}
l=\frac{1}{K} \left( e^{Kl}-1\right),
\end{equation}
i.e., to second order,
\begin{equation}
l=r-\frac{1}{2} K r^2.
\end{equation}
 In the first version of this paper, we based ourselves on the fact that we are now brought back to the RS \cite{RSG99} calculation, but with $K$ taking the value $\sqrt{\Lambda/3}$ instead of $H_0/c$. However, arrived at that point, RS \cite{RSG99} have identified $r$ with $c t$, which would yield $\delta l= (1/2) K c^2 t^2$, and therefore an acceleration $K c^2$. This is an incorrect identification, since the time variable one shoud take here is the signal propagation time on the light cone $t_0$ and not the Pioneer time $t$. We actually get $r \approx c t_0= v t$, so that the correct result is
\begin{equation}
\delta l=\frac{1}{2} K v^2 t^2,
\end{equation}
yielding an acceleration
\begin{equation}
a_P=K v^2.
\end{equation}
This value is reduced by a factor $v^2/c^2$ with respect to the previous expectation $K c^2$. 
In particular, the RS result (which would be valid only provided the expansion applies at the solar system scale) should be $a_P= H_0 v^2/c$ instead of $H_0 c$. The expected effect of the cosmological constant is therefore:
\begin{equation}
a_P=\sqrt{\frac{\Lambda}{3}} \; v^2 =\frac{v^2}{\sqrt{3} \,L_U}= \Omega_{\Lambda}^{1/2} H_0 \; \frac{v^2}{c}.
\end{equation}
In the case of the Pioneer spacecraft, $v\approx 12$ km/s, so that the reduction factor is $\approx 1.6 \times 10^{-9}$. The expected effect is finally very small, of the order of $10^{-16}$ cm.s$^{-2}$.

We conclude that the ``Machian" local effect of the cosmological constant (nor of the Hubble constant), due to the fact that usual local coordinates do not represent the true inertial frame, cannot explain an acceleration of the order of $\approx 10^{-7}$ cm.s$^{-2}$ as in the Pioneer effect. Note also that recently improved modelling of the reflective thermal acceleration of the Pioneer spacecraft has yielded a contribution of 25\% to 75\% to the total acceleration \cite{Francisco2011}. It is therefore not unlikely that the full effect be finally of thermal origin.

As concerns the effect theoretically established here, it is linked to the fact that our solar system does not achieve an inertial frame. The true inertial frame is actually relative to and determined by the global Universe, which manifests itself locally in terms of the horizon scale $\sqrt{3} \, L_{U}$. Is it possible to measure locally the cosmological constant  at the scale of the solar system by using this effect ?

An acceleration of the order of the observed Pioneer one, 
\begin{equation}
a_P=c^2 \sqrt{\Lambda/3}, 
\end{equation}
could be observed for a clock having a speed $v \approx c$ close to the speed of light. This effect amounts to 
\begin{equation}
a_P = (6.02 \pm 0.34)\times 10^{-8} \;{\rm  cm.s}^{-2}
\end{equation}
according to the WMAP 5 yrs results ($\Omega_\Lambda= 0.742 \pm 0.030$, $h = 0.719 \pm 0.026$) \cite{WMAP5yr}. A more precise expectation can be given in terms of a theoretically predicted value of the cosmological constant \cite{liwos,Nottale2011}, $\Lambda_{\rm th} = (1.3628 \pm 0.0003) \times 10^{-56}$ cm$^{-2}$. One gets in this case an acceleration
\begin{equation}
a_P = (6.0577\pm 0.0006)\times 10^{-8} \; {\rm  cm.s}^{-2}.
\end{equation}

 With a spacecraft moving at a speed of 1000 km/s, the effect would be $10^{-5}$ times the Pioneer effect (whose displacement has been $2.4\times 10^7$ m over seven years), i.e. a displacement of 240 m with respect to its trajectory without the additional acceleration. It is not impossible that such values could be reached in the framework of a dedicated space mission.

\end{document}